\documentclass{llncs}
\usepackage{makeidx}  
\usepackage{tabularx}
\usepackage{url}
\usepackage{multirow}
\usepackage[justification=centering]{caption}
\usepackage{booktabs}
\usepackage{graphicx}
\usepackage{algorithm}
\usepackage{algorithmic}
\usepackage{subfigure}
\usepackage{color}
\usepackage{amssymb,amsmath}
\usepackage{array,etoolbox}
\usepackage[flushleft]{threeparttable}
\preto\tabular{\setcounter{magicrownumbers}{0}}
\newcounter{magicrownumbers}

\begin{document}
\frontmatter          
\pagestyle{headings}  

\title{An Analysis of Privacy-Aware Personalization Signals by Using Online Evaluation Methods}
\author{Arjumand Younus\inst{1} \and M. Atif Qureshi\inst{2}}

\institute{Insight Centre for Data Analytics, \\
University College Dublin, Dublin, Ireland\\
\and
Centre for Applied Data Analytics Research, \\
University College Dublin, Dublin, Ireland\\
\email{\{arjumand.younus@ucd.ie,muhammad.qureshi@ucd.ie\}}}

\maketitle              

\begin{abstract}
Personalization despite being an effective solution to the problem information overload remains tricky on account of multiple dimensions to consider. Furthermore, the challenge of avoiding overdoing personalization involves estimation of a user's preferences in relation to different queries. This work is an attempt to make inferences about when personalization would be beneficial by relating observable user behavior to his/her social network usage patterns and user-generated content. User behavior on a search system is observed by means of team-draft interleaving whereby results from two retrieval functions are presented in an interleaved manner, and user clicks are utilised to infer preference for a certain retrieval function. This improves upon earlier work which had limited usefulness due to reliance on user survey results; our findings may aid real-time personalization in 
search systems by detecting a user-related and query-related personalization signals.

\end{abstract}

\section{Introduction}
In recent years personalized Web search has emerged as a promising way to improve the search quality through customization of search results for people with different information interests and goals \cite{sontag2012probabilistic,zamani2017situational}. Personalization however is a tricky endeavour due to the involvement of various factors mainly across the user dimension or the query dimension \cite{cai2017behavior,teevan2008personalize}. Of these, the user dimension in particular imposes privacy-related concerns adding a significant amount of complexity on account of users unwilling to share their search engine usage data \cite{wang2010exploring,younus2014investigation}. To further complicate matters, there is the challenge to avoid overdoing personalization, which may harm the user experience; so far work on achieving a balance between producing personalized and non-personalized results considers the query dimension or the user dimension in isolation with each other \cite{cai2017behavior,teevan2008personalize}. This paper argues for a framework that takes into consideration user-related personalization signals in combination with query-related personalization signals; we utilise characteristics of a user's social network usage data for these signals thereby alleviating the privacy considerations that arise due to personalization.

We conducted a similar study previously \cite{younus2014investigation,younus2013predictors} whereby a user survey was conducted to gather data about users' personalization preferences and their social network usage habits. However, the reliance on user survey responses limits the usefulness of those studies and in this work we attempt to make useful and reliable inferences through utilization of online evaluation methods. 

Online evaluation methods essentially comprise controlled experiments over a user population in order to reach a valid conclusion as to which system is preferred by the users. The two available choices are AB testing and interleaving. We utilize interleaving due the fact that it requires few interactions. To the best of our knowledge, no prior work utilizes online evaluation for detecting signals of personalization adjustments. Such signals can essentially aid the development of privacy-aware personalization that can adaptively adjust the level of personalization for different users. To the aim of identifying these signals, we conduct a user-study whereby users are presented with interleaved search results from two retrieval functions of which one is a personalized version. Within the study, we measure the degree to which personalization is preferred by means of a newly proposed ``Personalization Entropy" both by utilising observable clicks across real information needs, and by analysing correlations between various aspects of social network usage behavior/data and personalization preferences. Our results demonstrate that users who engage in high levels of communication over Twitter (in the form of mentions) prefer personalization to a larger degree. Furthermore, with respect to different topical aspects of a query the users prefer personalization for query topics across which their Twitter network contains information in the form of related keywords and keyphrases.

\section{Experimental Methodology}
\label{expmethod}
As the core of this work is investigating the link between social network usage patterns and personalization preferences, we focus on an efficient personalization mechanism that simply reranks the retrieved documents using a similarity measure between the returned document \textit{t} and a user \textit{u}'s profile \textit{p}\footnote{This is essentially on account of the need for methods that are fast enough to operate
in real-time which is crucial for online evaluation.}. Note that we utilise users' tweets for constructing their user profile in line with previous work \cite{younus2014language}.

The experimental procedure is setup as follows. Users were asked to deploy a browser plugin that detects Google Web searches and sends the search query and the unique user identifier to the server, which in turn fetches top 50 search results from Google. The plugin then presents interleaved results to the user whereby the original and personalized rankings are used to produce a combined ranking. We utilise the Team-Draft Interleaving (TDI) \cite{radlinski2008does} method on account of its tendency to minimize a bias towards rankings. A total of 137 users participated in our study who were asked to perform regular day-to-day searches with real information needs, and the study was conducted over a period of two months from 4th March, 2017 to 4th May, 2017.

In order to study the correlations between various aspects of social network usage and personalization potential, we have performed our study by analysing the following aspects of a user's microblog behavior:

\begin{itemize}
\item \textbf{Tweeting Behavior:} This aspect captures the amount of user engagement over Twitter in the form of number of tweets, mentions and retweets. We also include tweeting frequency as a mechanism to measure Twitter usage frequency \cite{younus2014investigation,younus2013predictors}.
\item \textbf{Twitter Network:} This aspect captures statistics about a user's Twitter network such as number of followers, number of Twitterers followed, number of Twitterers retweeted, number of Twitterers mentioned etc.
\item \textbf{Twitter Topical Densities:} This aspect captures statistics about topical content within a user's tweets such as keywords and their associated frequencies. Note that this is the aspect that captures both the user and query dimension, and is hence crucial to understanding a user's personalization preferences. 
\item \textbf{Twitter Network Topical Densities:} This aspect is similar to \textit{``Twitter Topical Densitiies"} with the difference that it takes into account topical content of a user's Twitter network such as those he follows, retweets, and mentions. 
\end{itemize}
More specifically, the topical densities mentioned above are computed using the approach proposed by Zhao et al. \cite{zhao2011topical} where we first extract topical keywords and topical keyphrases \footnote{We extract both unigram and bigram keywords for the purpose of topical densities' computation.} by combining tweets of the user under consideration and his/her Twitter network (i.e., those followed, retweeted and mentioned) and these are then used in relation to issued topics within user queries. A detailed description of how topical densities are computed from query topics follows below. To study the effect of social network usage on personalization preferences of users, we devise a measure that captures the personalization entropy measure inspired from the click entropy measure \cite{dou2007large}:

\begin{align}
PersonalizationEntropy( u_{qtopic} )= p (per_{u} \mid qtopic) *log \ p (per_{u} \mid qtopic) 
\end{align}

Here, $p (per_{u} \mid qtopic)$ represents the probability that user \textit{u} clicks a personalized URL query topic \textit{qtopic}. This measure essentially captures the amount of uncertainty associated with personalization preferences for a given user corresponding to various query topics: a lower personalization entropy value implies that many personalized urls were clicked whereas a higher personalization entropy value implies many non-personalized urls were clicked. Our measure is unique in that it models both the user and the query topic simultaneously. Note that our personalization entropy measure is particularly suited to online  evaluation methods and is helpful in the correlation analysis as outlined in Section \ref{expresults}. A variant of personalization entropy sums over all query topics of user \textit{u}, and thereby indicates his/her personalization entropy:

\begin{align}
PersonalizationEntropy( u )=\sum_{{qtopic}} p (per_{u} \mid qtopic) *log \ p (per_{u} \mid qtopic) 
\end{align}

We merge related queries into a single query topic by means of utilisation of Wikipedia categories using a variant of the approach by Gabrilovich and Markovitch \cite{gabrilovich2007computing}; through this method the queries ``brexit", ``theresa may europe plan", and ``scientific mobility europe and uk" will be classified as a single query topic. For computation of topical densities from Twitter profiles, we utilise the words (unigrams and bigrams) appearing within search result snippets and check for their occurence in the Twitter topical keywords; and density scores are then computed from keyword frequencies within Twitter-based profiles. As an example, consider the query topic ``brexit" with queries ``brexit", ``theresa may europe plan", and ``scientific mobility europe and uk" with the set \textit{W} representing words returned in search results and the set \textit{V} representing all the keywords and keyphrases in the Twitter profile of the considered user; a topical density score corresponding to this query topic is computed as follows:

\begin{align}
TopicalDensity_{qtopic}= \sum_{w \in  W} \dfrac{n(w,t)}{\sum_{w' \in  V}n(w',t)}
\end{align}
Here, $n(w,t)$ denotes the no. of tweets containing word `w'. Topical density scores essentially capture the normalized term frequencies for terms appearing in search results of a given query topic.

\section{Experimental Results and Discussion}
\label{expresults}
The analysed correlations between ``Tweeting Behavior" and ``personalization entropy" are shown in Table 1, wheareas the analysed correlations between ``Twitter Network" and ``personalization entropy" are shown in Table 2. Note that we apply a threshold to the number of tweets and to the number of Twitterers followed above which statistically significant correlations are observed;
here we report the results of the user-level personalization entropy by using the formula in Equation (2).

\begin{table}
 \begin{center}
  \begin{tabular}{|c|c|c|c|c|c|c|}
\cline{1-7}
Analysed & \multicolumn{6}{ |c| }{Dependent Variables} \\ \cline{2-7}
Correlations& \multicolumn{2}{|c|}{No. of Tweets} &\multicolumn{2}{|c|}{No. of Mentions} &\multicolumn{2}{|c|}{No. of ReTweets}  \\ \hline
\textit{Pearson}& \multicolumn{2}{|c|}{0.464} & \multicolumn{2}{|c|}{-0.822**} &\multicolumn{2}{|c|}{0.261}     \\ \cline{2-7} \hline
\textit{Spearman}& \multicolumn{2}{|c|}	{0.367} & \multicolumn{2}{|c|}{-0.876*} &\multicolumn{2}{|c|}{0.341}     \\ \cline{2-7} \hline
\textit{Regression Analysis}& \multicolumn{2}{|c|}{0.612} & \multicolumn{2}{|c|}{-0.724***} &\multicolumn{2}{|c|}{0.387}     \\ \cline{2-7} \hline
\end{tabular}
\end{center}
\begin{tablenotes}
      \small
      \item Note *p$<$.05, **p$<$.01, ***p$<$.001
    \end{tablenotes} 
  \caption{Analysis of Correlations between ``Tweeting Behavior" Features and ``Personalization Entropy"}
\end{table}

\begin{table}
 \begin{center}
  \begin{tabular}{|c|c|c|c|c|c|c|c|c|}
\cline{1-9}
Analysed & \multicolumn{8}{ |c| }{Dependent Variables} \\ \cline{2-9}
Correlations& \multicolumn{2}{|c|}{No. of Followers} &\multicolumn{2}{|c|}{No. of Followed} &\multicolumn{2}{|c|}{No. of Mentioned} &\multicolumn{2}{|c|}{No. of Retweeted}   \\ \hline
\textit{Pearson}& \multicolumn{2}{|c|}{0.451} & \multicolumn{2}{|c|}{0.567} &\multicolumn{2}{|c|}{-0.834**}&\multicolumn{2}{|c|}{0.461}     \\ \cline{2-9} \hline
\textit{Spearman}& \multicolumn{2}{|c|}{0.345} & \multicolumn{2}{|c|}{0.472} &\multicolumn{2}{|c|}{-0.921***}&\multicolumn{2}{|c|}{0.493}     \\ \cline{2-9} \hline
\textit{Regression Analysis}& \multicolumn{2}{|c|}{0.612} & \multicolumn{2}{|c|}{0.556} &\multicolumn{2}{|c|}{-0.864**}&\multicolumn{2}{|c|}{0.513}     \\ \cline{2-9} \hline
\end{tabular}
\end{center}
\begin{tablenotes}
      \small
      \item Note *p$<$.05, **p$<$.01, ***p$<$.001
    \end{tablenotes} 
  \caption{Analysis of Correlations between ``Twitter Network" Features and ``Personalization Entropy"}
\end{table}

Table 3 on the other hand shows the correlations between ``Personalization Entropy" measure across a user and across different query topics (shown in Equation (1) of Section \ref{expmethod}), and topical densities from within query topics.

\begin{table}
 \begin{center}
  \begin{tabular}{|c|c|c|c|c|c|c|c|c|}
\cline{1-9}
Analysed & \multicolumn{8}{ |c| }{Dependent Variables} \\ \cline{2-9}
Correlations& \multicolumn{2}{|c|}{Topical Density$_{User}$} &\multicolumn{2}{|c|}{Topical Density$_{F}$} &\multicolumn{2}{|c|}{Topical Density$_{M}$} &\multicolumn{2}{|c|}{Topical Density$_{R}$}   \\ \hline
\textit{Pearson}& \multicolumn{2}{|c|}{-0.751***} & \multicolumn{2}{|c|}{-0.567} &\multicolumn{2}{|c|}{-0.876**}&\multicolumn{2}{|c|}{-0.9214**}     \\ \cline{2-9} \hline
\textit{Spearman}& \multicolumn{2}{|c|}{-0.658*} & \multicolumn{2}{|c|}{-0.431**} &\multicolumn{2}{|c|}{-0.913***}&\multicolumn{2}{|c|}{-0.721**}     \\ \cline{2-9} \hline
\textit{Regression Analysis}& \multicolumn{2}{|c|}{-0.612**} & \multicolumn{2}{|c|}{-0.556*} &\multicolumn{2}{|c|}{-0.775***}&\multicolumn{2}{|c|}{-0.822***}     \\ \cline{2-9} \hline
\end{tabular}
\end{center}
\begin{tablenotes}
      \small
      \item Note *p$<$.05, **p$<$.01, ***p$<$.001
    \end{tablenotes} 
  \caption{Analysis of Correlations between ``Twitter Topical Densities" Features and ``Personalization Entropy"}
\end{table}

According to Tables 1 and 2, the number of mentions and the number of Twitterers mentioned have a high, negative correlation with personalization entropy thereby implying that users with a high volume of mentions and a high number of Twitter users mentioned prefer personalized search results. Intuitively, mentions account for highly specific conversations and represent specific user interests. Note that this is in contrast to our previous findings where mentions were found to be negatively correlated with a preference for personalization \cite{younus2014investigation,younus2013predictors}; and this further proves the need for a more reliable means of observing user search behavior instead of reliance on user survey results.

Finally, Table 3 depicts a significant outcome of a user's Twitter content being highly reflective of his/her information-seeking needs, and personalization entropy is low (meaning high preference for personalized urls). Moreover, topical densities of a user's Twitter network showing a negative, high correlation with personalization entropy depicts the significance of a user's associations with a query topic in the form of his/her network and/or his/her content. This association may help in inference of a personalization signal in real-time search systems and this could alleviate the need for utilising private user content such as desktop data, search history etc.

%

\vspace{0.3em} 

%
\bibliographystyle{abbrv}
\bibliography{sigproc}

\vfill

%
%
\end{document}